\documentclass[%
 reprint,
superscriptaddress, 
 amsmath,amssymb,
 aps,
 prmaterials
]{revtex4-2}

\usepackage{graphicx}
\usepackage{dcolumn}
\usepackage{bm}
\usepackage{siunitx}

\usepackage[dvipsnames]{xcolor}

\newcommand{\MnGe}{\ensuremath{\text{Mn}_3\text{Ge}\ }}%
\newcommand{\hexagonal}{hexagonal $\text{D}\text{0}_{19}\ $}%
\newcommand{\tetragonal}{tetragonal $\text{D}\text{0}_{22}\ $}%
\begin{document}

\preprint{APS/123-QED}

\title{Optimised growth of epitaxial $\mathrm{D0_{19}}$ $\mathrm{Mn_{3}Ge}(0001)$ thin films via single target sputtering}

\author{Servet Ozdemir}
\email{s.ozdemir@leeds.ac.uk}
\affiliation{School of Physics and Astronomy, University of Leeds, LS2 9JT, Leeds, UK}

\author{Ping-Luen Ho}
\affiliation{Department of Materials, Imperial College London, London, SW7 2AZ, UK}
\author{Shelly Conroy}
\affiliation{Department of Materials, Imperial College London, London, SW7 2AZ, UK}

\author{Callum Brennan-Rich}
\affiliation{School of Physics and Astronomy, University of Leeds, LS2 9JT, Leeds, UK}

\author{Matthew Rogers}
\affiliation{School of Physics and Astronomy, University of Leeds, LS2 9JT, Leeds, UK}

\author{Philippa Shepley}
\author{Ahmet Yagmur}
\author{Ismet Gelen}
\author{Sehwan Song}
\affiliation{School of Physics and Astronomy, University of Leeds, LS2 9JT, Leeds, UK}

\author{Zabeada Aslam}
\affiliation{School of Chemical and Process Engineering, University of Leeds, LS2 9JT, Leeds, UK}
\author{Mannan Ali}
\affiliation{School of Physics and Astronomy, University of Leeds, LS2 9JT, Leeds, UK}

\author{Sandrine Heutz}
\affiliation{Department of Materials, Imperial College London, London, SW7 2AZ, UK}

\author{Gavin Burnell}
\email{G.Burnell@leeds.ac.uk}
\author{B J Hickey}
\email{B.J.Hickey@leeds.ac.uk}
\affiliation{School of Physics and Astronomy, University of Leeds, LS2 9JT, Leeds, UK}

\date{\today}

\begin{abstract}
Materials with chemical identity $\text{Mn}_3\text{X}$, where X is Sn, Ga, or Ge, are Heusler alloys with intriguing properties that are fundamentally and technologically interesting. They have topological properties with a non-trivial band structure that identify them as Weyl semimetals in the bulk.  Their magnetic properties, which give rise to the topological band structure through spin-orbit coupling, are determined by their Kagome lattice structure, making them chiral non-collinear antiferromagnets.  Among these materials, $\text{Mn}_3\text{Ge}$ is particularly interesting because it has a larger Berry curvature, leading to a large anomalous Hall conductivity and motivating potential spintronic applications.  The desirable properties are only obtained when $\text{Mn}_3\text{Ge}$ has the $\text{D}\text{0}_{19}$ crystal structure, so producing an easily reproducible growth recipe is the main objective of this paper.  Here we show that ultra-high-quality thin films of $\text{Mn}_3\text{Ge}$ with $\text{D}\text{0}_{19}$ symmetry can be grown from a $\text{Mn}_5\text{Ge}$ target on an $\text{Al}_2\text{O}_3(0001)$ substrate with a Ru(0001) seed layer.  The expected non-collinear antiferromagnetic order was confirmed by the finding of weak zero-field magnetisation $\mathrm{(0.0086 \pm 0.0002) \,\mu_{B}/Mn}$ that does not saturate at high magnetic fields and has a Néel temperature of $\mathrm{(291 \pm 2)\, K}$.

\end{abstract}

\maketitle


\section{\label{sec:introduction}INTRODUCTION}

The observations of a large anomalous Hall effect (AHE) despite a vanishingly small magnetisation in Kagome crystals that display non-collinear antiferromagnetism \cite{Nakatsuji2015}, including \hexagonal \MnGe \cite{Kiyohara2016, Nayak2016}, have led to the birth of non-collinear antiferromagnetic spintronics \cite{Rimmler2024}. First-principles calculations have attributed AHE to a larger Berry curvature in the vicinity of the Fermi level \cite{Zhang2017}, which in the bulk leads to the maximum number of source and sink Weyl points \cite{Yang2017}. Progress has been made in thin film growth of $\text{Mn}_3\text{Ge}(0001)$ films through co-sputtering utilising an Ru(0001) seed layer \cite{Hong2020, Jeon2021}, where an Mn:Ge flux ratio of 5:1 has been identified as optimal \cite{Hong2020}. Using \MnGe single targets, both \tetragonal and $(20\bar{2}0)$ textured \hexagonal phases have been reported on STO(001) substrates \cite{Kurt2012,Wang2021_1} and \tetragonal phases on Ru(101) seed layers \cite{Kobayashi2020}. Single $\text{Mn}_5\text{Ge}$ targets, on the other hand, were found to lead to mixed-phase and/or mixed-orientation thin films on STO(001) and $\text{Al}_2\text{O}_3(1\bar{1}02)$ substrates \cite{Wang2022, Wang2025} despite being suggested to be optimal for \hexagonal phase by co-sputtering studies \cite{Hong2020}.

Here, utilising $\text{Mn}_5\text{Ge}$ targets and Ru(0001) seed layers, we demonstrate an initial growth of entirely tetragonal $\mathrm{{D0}_{22}}$, (112) textured films. Contrary to experiments in STO(001) films \cite{Wang2021_2}, by annealing \tetragonal(112) textured films at \SI{500}{\celsius} we show that they undergo a structural transition into \hexagonal (0001) films. We characterised the grown thin films through X-ray Diffraction (XRD), X-ray Reflectivity (XRR), Reflection High Energy Electron Diffraction (RHEED), 4D-Scanning Transmission Electron Microscopy (4D-STEM) as well as Energy Dispersive X-ray Spectroscopy (EDX) and show the \hexagonal films in particular to have an epitaxial nature. Through magnetometry experiments, we confirm the two different magnetic characteristics the different phases are expected to show, \hexagonal phase being a non-collinear antiferromagnet \cite{Nagamiya1982, Tomiyoshi1983} and \tetragonal phase being a c-axis ferrimagnet \cite{Kadar1971}.

\begin{figure*}[htbp!]
\includegraphics[width=\textwidth]{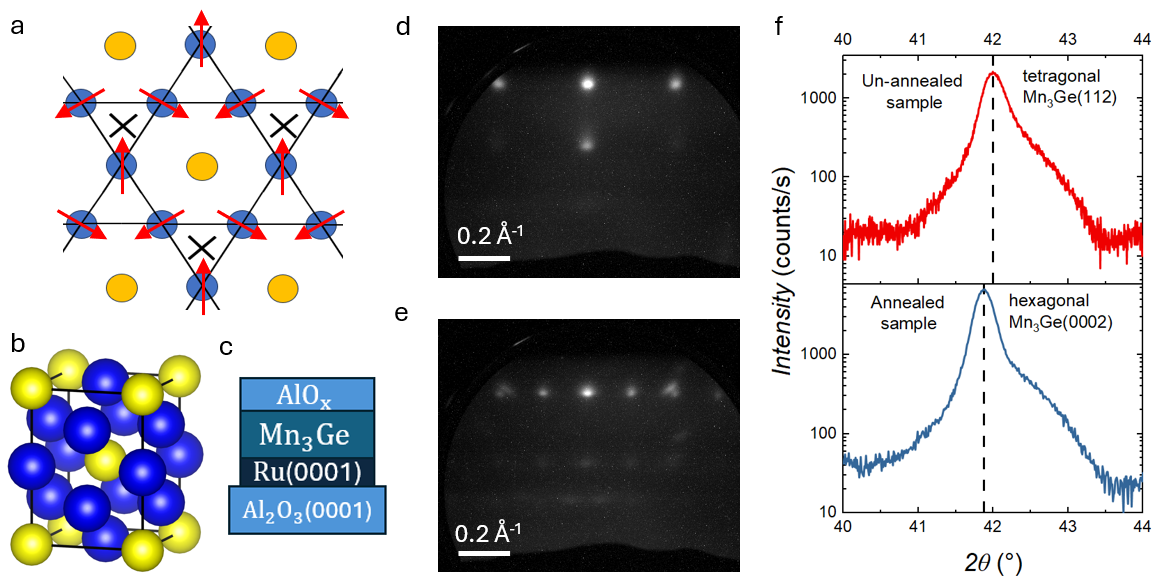}
\caption{\label{fig:structure} (a) Kagome crystal structure of (0001) orientated \hexagonal \MnGe\  film, with Mn atoms in blue and Ge atoms in yellow where the non-collinear ground state configuration of the spins are depicted with red arrows. (b) Unit cell of a c-axis elongated \tetragonal \MnGe\  structure with Mn atoms in blue and Ge atoms in yellow. (c) Schematic illustrating the heterostructure consisting of Ru/$\mathrm{Mn_{3}Ge}$/$\text{Al}\text{O}_\text{X}$ layers grown on c-plane sapphire where the thickness of the Ru seed layer is 7 nm and the $\text{Al}\text{O}_\text{X}$ cap layer is 5 nm. (d) RHEED image of the as-grown $\sim$ 40 nm \MnGe\  film at \SI{250}{\celsius}, for which, assuming a $2 \times 1$ super-lattice formation on the surface, lattice parameters of $a$ = $b$ =  $\mathrm{(3.78\pm0.01)~\text{\AA}}$ could be determined. (e) RHEED image obtained on the $\sim$ 40 nm \MnGe  film after annealing at \SI{500}{\celsius} for 1 hour, with additional diffraction spots emerging, suggesting a phase transition and lattice parameters of $a$ = $b$ = $\mathrm{(5.36\pm0.01)~\text{\AA}}$. (f) XRD peaks measured on un-annealed sample (upper panel - red) and annealed sample (lower panel - blue) showing a shift in measured peak position, suggesting a change in out-of-plane lattice parameter, from $\mathrm{(7.57\pm0.01) \,{\AA}}$ for $\text{D}\text{0}_{22}$ $\text{Mn}_3\text{Ge}$(112) peak to $\mathrm{(4.30\pm0.01)\, {\AA}}$ for $\text{D}\text{0}_{19}$ $\text{Mn}_3\text{Ge}$(0002) peak due to annealing induced structural transition. The shoulders are due to Ru(0002) peak in close proximity to \MnGe peaks. 
}
\end{figure*}

\section{\label{sec:experimental}METHODS}
Thin film growth by DC magnetron sputtering was carried out in a chamber with a base pressure of $\sim$ $\text10^{-10}$ mbar in an Ar pressure of $4.6 \times 10^{-3}$ mbar.
The growth of \MnGe films was carried out using Ru(0001) seed layers grown on c-plane sapphire followed by \MnGe from a $\text{Mn}_5\text{Ge}$ target. A Mn:Ge flux ratio of 5 to 1 has been previously demonstrated to be necessary \cite{Hong2020} to achieve the required stoichiometry. In order to obtain Ru(0001) layers, DC sputtering of Ru was carried out at a nominal substrate temperature of \SI{450}{\celsius} with a power of $\mathrm{25\, W}$ and a growth rate of $\approx$ $\mathrm{0.25 \,{\AA} s^{-1}}$. The substrate was then cooled to \SI{250}{\celsius} in order to obtain continuous layers of \tetragonal  \MnGe with (112) texture which were grown at a rate of $\approx$ $\mathrm{0.30 \,{\AA} s^{-1}}$ with a power of $\mathrm{20\, W}$. We annealed the films for 1 hour at \SI{500}{\celsius} to obtain \hexagonal \MnGe films with (0001) texture. Films were finally capped with a $\sim$ 5 nm alumina layer through RF sputtering from an $\mathrm{Al_2O_3}$ target. RHEED characterisation was carried out with STAIB Instruments system consisting of an electron gun which generates a 15 keV electron beam incident at a glancing angle less than \SI{5}{\degree} on the sample surface. XRR, XRD and  $\phi$-scans were carried out on a Bruker D8 X-ray diffractometer with a Cu source. For transmission electron microscopy characterisation of thin film structures cross-section lamellae were prepared using a dual beam focused Ga ion beam integrated scanning electron microscope (Thermo-Fisher Scientific FEI Helios 5). After electron and ion beam deposition of C and Pt on the surface, the samples were thinned to electron transparency. Transmission Electron Microscopy (TEM) and High-angle annular dark-field scanning TEM (HAADF-STEM) was performed using a probe-aberration-corrected Spectra 300 microscope (Thermo Fisher Scientific) at 300kV. 4D-STEM was collected via a Tescan Tensor Analytical STEM at 100 kV and a current of 250 pA. 4D-STEM was recorded with a probe size of 0.4 nm and a convergence of 6 mrad. The electron diffraction patterns obtained with 4D-STEM were later simulated using the Pyxem extension of the Python Hyperspy package \cite{pyxemorientationmapping2022}. All scripts used in the analysis are publicly available.  Magnetometry measurements were carried out using a SQUID-VSM Magnetometer (MPMS3 from Quantum Design) with a resolution of $10^{-8}$ emu and temperature control from \SI{400}-\SI{1.6}{K} with applied magnetic fields, $\mu_{0}H$ up to \SI{7}{T}.

\begin{figure*}[!t]
\centering
\includegraphics[width=0.99\textwidth]{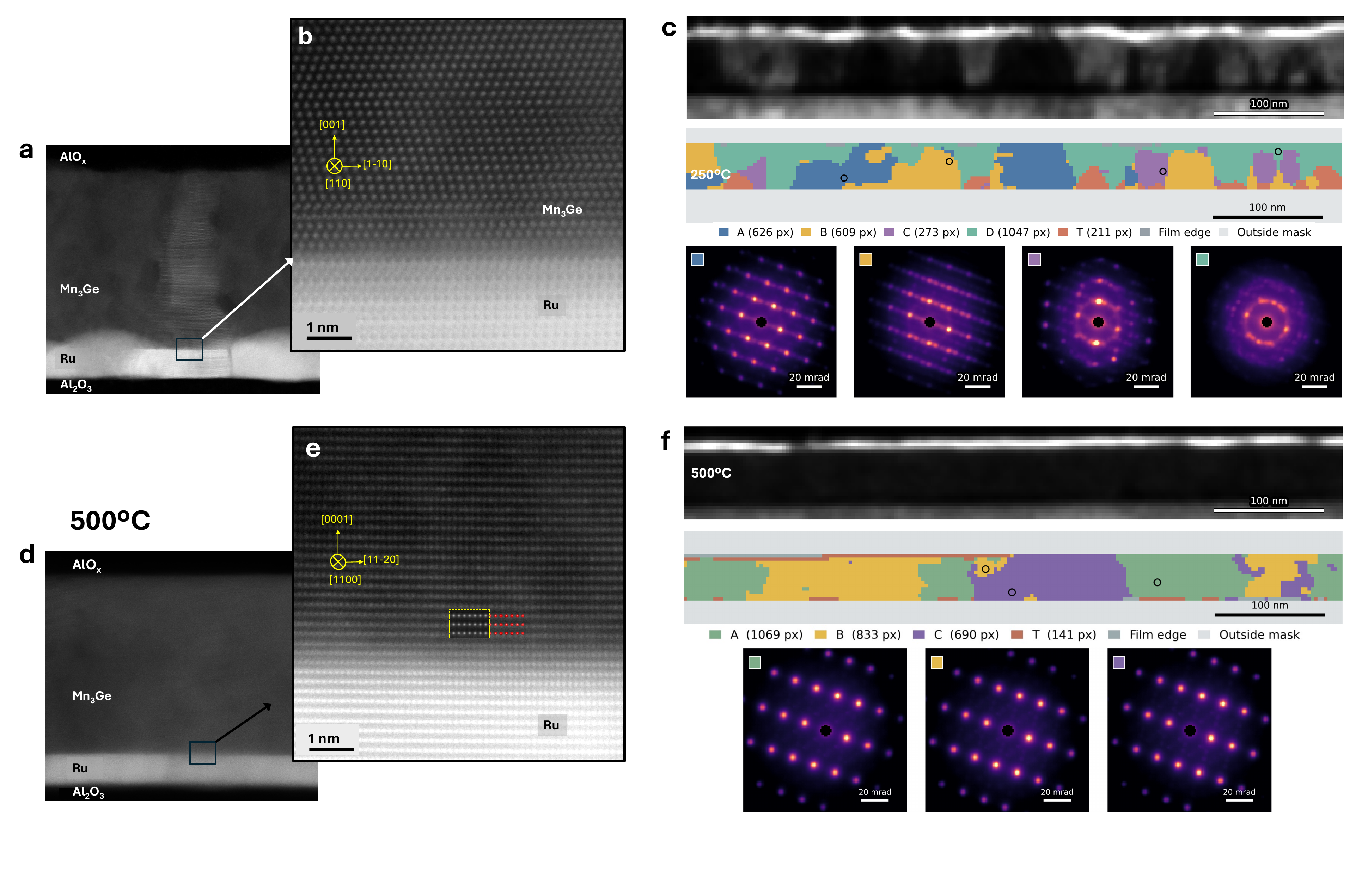}
\caption{\label{fig:TEM}
(a) Low-magnification HAADF-STEM image of the as-grown Mn$_3$Ge film deposited at \SI{250}{\celsius}.
(b) Atomic-resolution HAADF-STEM image of the as-grown film.
(c) 4D-STEM diffraction classification of the as-grown film.
(d) Low-magnification HAADF-STEM image following annealing at \SI{500}{\celsius}.
(e) Atomic-resolution HAADF-STEM image of the annealed film.
(f) 4D-STEM diffraction classification of the annealed film.
}
\end{figure*}

\section{\label{sec:results}RESULTS AND DISCUSSION}

\subsection{\label{sec:structural}Structural transition}

The Kagome crystal structure as well as the ground state magnetic structure \cite{Soh2020} of the (0001) plane of topologically non-trivial \hexagonal $\text{Mn}_3\text{Ge}$  \cite{Nagamiya1982} is depicted in Figure \ref{fig:structure}a with the unit cell for the c-axis-elongated \tetragonal phase \cite{Kurt2012} shown in Figure \ref{fig:structure}b. The cross-section schematic of the Ru-seeded heterostructure with $\text{Al}_2\text{O}_3$(0001) substrate and $\text{Al}\text{O}_\text{X}$ cap layer is depicted in Figure \ref{fig:structure}c. Figure \ref{fig:structure}d shows RHEED characterisation of continuous \MnGe films grown at \SI{250}{\celsius}, with the diffraction spots yielding a lattice spacing of (2.67±0.01) {\AA}. Figure \ref{fig:structure}e shows RHEED characterisation post 1 hour annealing at \SI{500}{\celsius}, where a structural transition is evidenced, with RHEED data suggesting a lattice spacing of (5.36±0.01) \AA. Figure \ref{fig:structure}f shows XRD characterisation of un-annealed and annealed samples, where a shift in XRD peak is observed from $(42.00 \pm \num{0.01})\unit{\degree}$ to (41.90 ± \num{0.01})\unit{\degree} post annealing. See  \cite{supp}, Figure S1, for full XRD curves showing absence of any other peaks corresponding to other phases \cite{Wang2025}. The peaks can be indexed as (112) for the tetragonal structure as grown at \SI{250}{\celsius}, and (0002) for hexagonal phase obtained post annealing at \SI{500}{\celsius}. For the (0001) textured \hexagonal phase, the RHEED lattice spacing of (5.36±0.01) \AA, corresponds to in-plane lattice parameters $a$ and $b$. The c-axis is confirmed to be (4.30±0.01) \AA, by XRD. A fitting of the XRD peak for the annealed structure is shown in \cite{supp} where an out-of-plane grain size of (40.4±0.2) nm is estimated which is showing that the crystal is coherent over the entire thickness of the thin film. We have also carried out further XRD  $\phi$-scan measurements on the annealed sample \cite{supp}, to confirm the hexagonal order with extracted lattice parameters. The $\phi$-scan shows six peaks in both Ru seed layers and $\text{Mn}_3\text{Ge}$  layers in agreement with the 6-fold symmetry of the hexagonal crystals. The expected twinning effects associated with columnar growth of $\text{Mn}_3\text{Ge}$ films are present, yet only 6 prominent peaks appear, suggesting that the domains associated with columnar growth have a phase relationship that enables the onset of bulk-like long-range magnetic order. For the RHEED analysis of the un-annealed \tetragonal sample a (112) surface azimuth of [$\bar{1}$10] could be defined for the RHEED beam, which would yield the lattice spacing $d_{[\bar{1}10]}=a\sqrt{2}$ as $a=b$. Assuming a $2 \times 1$ superlattice formation on the surface, one can determine the lattice parameters $a$ and $b$ as (3.78±0.01) \AA, in agreement with the values in the literature \cite{Kurt2012, Wang2021_2}. Having obtained $a$ and $b$ lattice parameters we can use the XRD measurement to obtain the $c$ lattice parameter. By using the measured 2$\theta$ value, (42.00 ± \num{0.01})\unit{\degree}, we can calculate the lattice spacing $d_{112}$ as (2.15±0.01) \AA, which would yield $c$ lattice parameter of (7.22±0.01) \AA, which is also in agreement values in the literature \cite{Kurt2012, Wang2021_2}.  
\begin{figure*}[htbp!]
\includegraphics[scale=0.7]{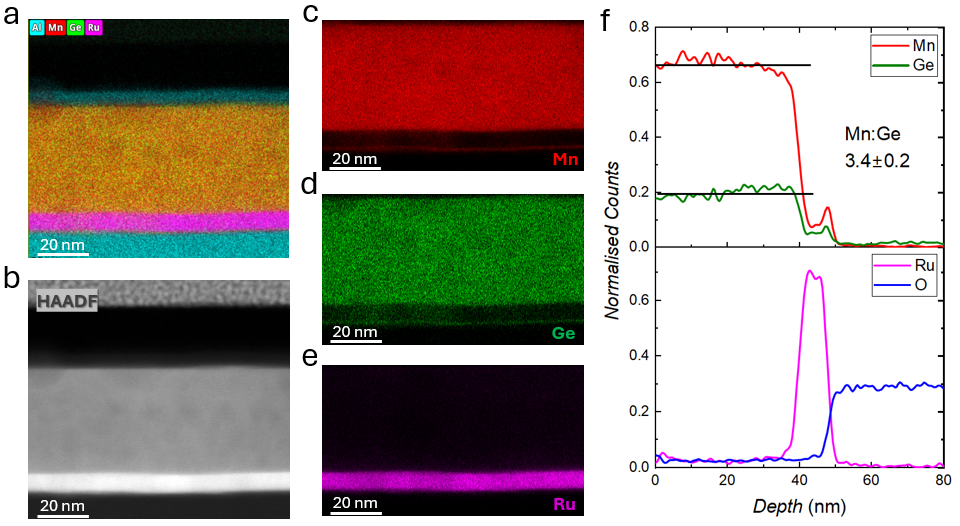}
\caption{\label{fig:EDX}  (a) Multi-element STEM Energy Dispersive X-ray Spectroscopy (EDX) map and (b) corresponding HAADF image on a 40 nm, annealed D0$_{19}$ Mn$_{3}$Ge sample. (c) (d) (e) Single element maps (red - Mn, green -Ge, lilac-Ru) corresponding to the multi-element EDX map in (a). (f) A Mn:Ge stoichiometric ratio of 3.4±0.2 can be determined from the EDX elemental line cuts, and the effectiveness of AlO$_{X}$ capping in protecting the film from oxidation can be seen.}
\end{figure*}

\subsection{\label{sec:TEM}TEM characterisation}

Following the structural characterisation by XRD and RHEED, scanning transmission electron microscopy (STEM) was used to investigate the local crystal structure and microstructure of the Mn$_3$Ge films before and after annealing (Figure~\ref{fig:TEM}). The as-grown film deposited at \SI{250}{\celsius} exhibits a relatively rough top interface and a heterogeneous microstructure (Figure~\ref{fig:TEM}a), with atomic-resolution HAADF-STEM imaging consistent with the tetragonal D0$_{22}$ structure (Figure~\ref{fig:TEM}b). Following annealing at \SI{500}{\celsius}, the film exhibits a smoother top interface (Figure~\ref{fig:TEM}d), while atomic-resolution imaging reveals the hexagonal D0$_{19}$ structure (Figure~\ref{fig:TEM}e). Additional HAADF-STEM and LAADF-STEM images of the annealed film are presented in Supplementary Figure S4. We acquired the atomic-resolution images along different crystallographic projections due to the orientations of the FIB-prepared lamellae, but both are consistent with the structural transformation identified by XRD and RHEED.

To investigate the spatial variation in crystal structure across the films, 4D-STEM datasets were acquired and analysed using the open-source Python package HyperSpy~\cite{deLaPena2017}. Virtual bright-field images were reconstructed from the diffraction datasets (Figure~\ref{fig:TEM}c,f), and unsupervised classification was used to group diffraction patterns with similar crystallographic characteristics and map their spatial distribution. The as-grown film contains numerous smaller regions with varying diffraction characteristics (Figure~\ref{fig:TEM}c), consistent with its heterogeneous microstructure and the polycrystalline regions observed by conventional TEM (Supplementary Figures S5 and S6, \cite{supp}). In comparison, the annealed film exhibits larger, more spatially continuous regions with similar diffraction patterns (Figure~\ref{fig:TEM}f), suggesting grain growth and improved crystallographic uniformity following annealing. Further analysis of the diffraction-pattern correlations, projected class occupancies and reflection intensities is presented in Supplementary Figures S7 and S8, \cite{supp}. These classifications identify regions with similar diffraction characteristics rather than directly measuring individual grains.

Comparison of the experimental diffraction patterns with simulations based on lattice parameters obtained from RHEED and XRD further supports the transformation from tetragonal D0$_{22}$ Mn$_3$Ge to hexagonal D0$_{19}$ Mn$_3$Ge following annealing (Supplementary Figure S9, \cite{supp}). Additional reflections are observed in some regions of the annealed film (Supplementary Figure S10, \cite{supp}), which may be associated with the formation of an incommensurate superstructure at the lamella surface~\cite{Hadermann2012}.

Finally, X-ray reflectivity (XRR) measurements were used to examine changes in film thickness and interfacial roughness following annealing. The Mn$_3$Ge thickness remains largely unchanged, decreasing slightly from $32.1 \pm 0.5$ nm in the as-grown film (Supplementary Figure S11a) to $31.5 \pm 0.2$ nm after annealing at \SI{500}{\celsius} (Supplementary Figure S11b). However, the fitted Mn$_3$Ge root-mean-square roughness decreases from $1.41 \pm 0.07$ nm to $0.92 \pm 0.05$ nm post-annealing, consistent with the smoothening of the Mn$_3$Ge interfaces as suggested by HAADF-STEM cross-sectional images. Further corresponding XRR fitting parameters are provided in Supplementary Information, \cite{supp}, Tables S1 and S2.

\begin{figure}[htbp!]
\includegraphics[scale=0.325]{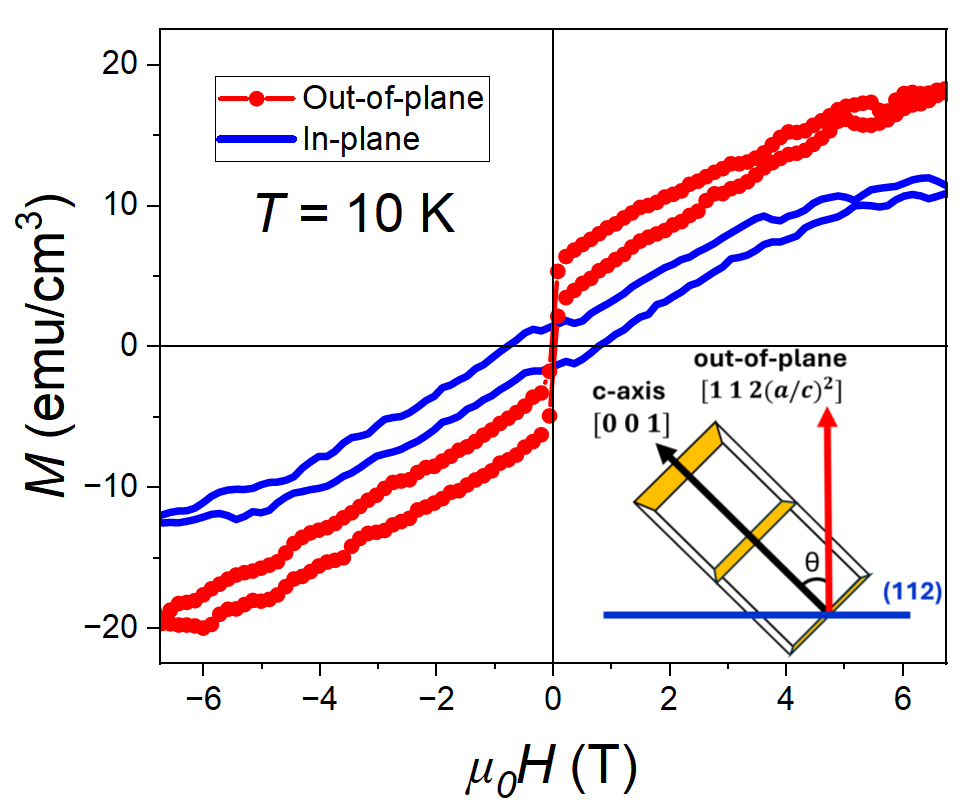}
\caption{\label{fig:MHloop} Magnetisation loops measured on the \tetragonal \MnGe(112) 40 nm sample for in-plane (blue) and out-of-plane (red) orientations of magnetic field at \SI{10}{K}. 
}
\end{figure}
\begin{figure*}[htbp!]
\includegraphics[scale =0.75]{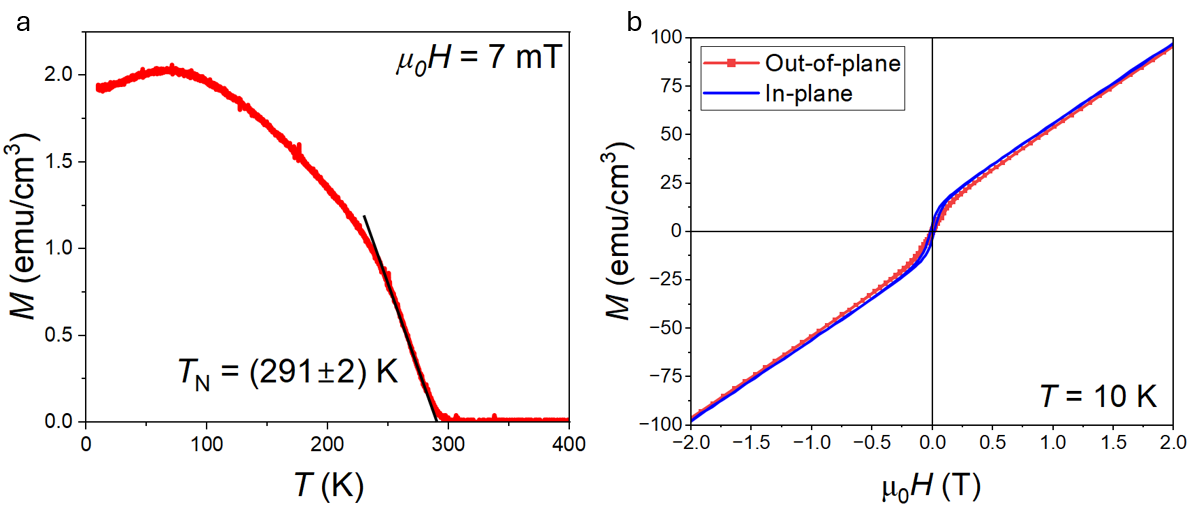}
\caption{\label{fig:magnetometry} a) Magnetisation versus temperature curves measured at $\mu_0 H = 7$ mT for \hexagonal \MnGe film of 40 nm thickness. b) Magnetisation loops measured for \hexagonal \MnGe(0001) (40 nm) sample for in plane (blue) and out-of-plane (red) orientations of magnetic field at \textit{T}= 10 K.
}
\end{figure*}
\subsection{\label{sec:magnetometry}Energy Dispersive X-ray Spectroscopy}
Uncapped \hexagonal \MnGe films have been shown to form a passivating oxidised layer $\mathrm{Mn_{3}GeO_{X}}$ of $\approx$ $\mathrm{8\, nm}$ thickness \cite{Hong2020}. The use of an $\text{Mn}_5\text{Ge}$ single target for growth, also raises the question of the  Mn:Ge stoichiometry of the grown structure. In order to determine both the effectiveness of the $\text{AlO}_\text{X}$ capping layer and the stoichiometric Mn:Ge ratio, we have carried out STEM-EDX as shown in Figure \ref{fig:EDX}. A multi-element map shown in Figure \ref{fig:EDX}a and a corresponding HAADF image as well as Mn, Ge and Ru single element maps are shown in Figure \ref{fig:EDX}b, c d and e. The Mn and Ge elemental maps suggest an interdiffusion of small amounts of both elements below the Ru layer post-annealing, which is also confirmed by the line profiles shown in Figure \ref{fig:EDX}f, upper panel. From these curves, it is possible to estimate the Mn:Ge stochiometric ratio, which is found to be $\mathrm{3.4 \pm 0.2}$. This agrees with other studies, including bulk systems, where an excess of Mn is necessary to stabilise the \hexagonal phase \cite{Kiyohara2016, Hong2020}. Lastly, the line profile for O is plotted on Figure 3f, bottom panel. It can be seen that oxygen is undetectable within the EDX profile until the Ru seed layer (bottom panel lilac curve) Al$_2$O$_3$ substrate inteface, showing effectiveness of the AlO$_X$ cap in protecting the sample from oxidation.

\subsection{\label{sec:magnetometry}Magnetometry}

The \hexagonal and \tetragonal phases differ in their magnetic properties. The \tetragonal films are c-axis ferrimagnets in bulk \cite{Kadar1971}, and thin film studies on (001) orientated films have demonstrated the c-axis as their easy axis in agreement with bulk \cite{Kurt2012, Wang2021_2}. In ultra-thin (001) textured films, a smaller coercivity has been observed \cite{Garg2025}. In our films, which have (112) texture (as well as significant polycrystalline regions, see \cite{supp}), the easy axis is therefore expected to be tilted $58.4 \unit{\degree}$ from the normal, see Supplemental Material, \cite{supp}, Supplemental Note 1. Figure \ref{fig:MHloop} shows out-of-plane (red) and in-plane (blue) hysteresis loops for the 40 nm (112) textured tetragonal $\text{D}\text{0}_{22}$ film. Neither the out-of-plane nor in-plane magnetic hysteresis loops show any saturation effects up to an applied magnetic field of 7 T. This is expected because the easy axis is tilted relative to the crystallite orientations, with the inset in Figure 4 showing the relative orientations between the easy axis (black arrow) and the out-of-plane magnetic field (red arrow), as well as the (112) sample plane (blue line) relative to a unit cell. It can be seen that, under an out-of-plane magnetic field (red curve), the sample exhibits a magnetisation component with low coercivity and therefore a soft-magnetic nature. We attribute this effect to polycrystalline regions in the film, as both bulk polycrystalline films \cite{Yamada1990} and thin films within the ultra-thin limit \cite{Garg2025} where crystallite sizes are comparable to the polycrystaline regions, have been found to show small coercivities. The in-plane loop, on the other hand, shows a larger coercivity on the order of 0.75 T, in agreement with the extremely hard magnetic nature of bulk-like thin films when a magnetic field is applied along the c-axis \cite{Kurt2012, Wang2021_2}. We attribute the measured smaller net magnetisation for both in-plane and out-of-plane magnetic field configurations compared to (001) textured thin films \cite{Kurt2012} and the smaller coercivity for the in-plane field (0.75 T compared to 2.25 T \cite{Kurt2012}) to excess manganese atoms; in other words, the 3.4:1 stoichiometry for Mn and Ge atoms. The effect of excess Mn-induced reduction in net ferrimagnetic signal has previously been discussed \cite{Kurt2012, Yamada1990}.

The hexagonal, topologically non-trivial $\text{D}\text{0}_{19}$ \MnGe\  films are found to show a weak magnetic signal that is a result of the non-collinear antiferromagnetic order \cite{Ohoyama1961, Yamada1988, Nagamiya1982}. Figure \ref{fig:magnetometry}a shows magnetisation versus temperature data measured with magnetic field of 7 mT. The weak magnetic signal due to the onset of non-collinear antiferromagnetic order is clearly visible below the Néel temperature $\textit{T}_{N}$, which can be determined from the tangent to the temperature-dependent magnetisation, as shown with the black line. This extracted Néel temperature of $\mathrm{(291 \pm 2)\,K}$ is smaller than the bulk temperature of $\mathrm{360 \,K}$ \cite{Yamada1988}, which would be an expected consequence of reduced thickness of the thin-film.  Magnetic hysteresis loops of bulk $\text{Mn}_3\text{Ge}$ crystals, as well as related systems $\text{Mn}_3\text{Sn}$ and $\text{Mn}_3\text{Ga}$, have been found to exhibit linear magnetisation growth as a function of magnetic field as well as a low magnetic field offset \cite{Nayak2016, Nakatsuji2015, Song2024}. For thin films, however, magnetic hysteresis loops have been reported to show magnetisation saturation \cite{Hong2020, Markou2018, isis2026}, which strongly disagrees with theoretical models \cite{Liu2017} and the bulk signatures of the expected non-collinear antiferromagnetic nature of the films. After carefully subtracting the diamagnetic substrate contribution, we find that these magnetic saturation effects are absent in our ultra-high-quality $\text{Mn}_3\text{Ge}$(0001) films, as shown in Figure \ref{fig:magnetometry} (see \cite{supp}, Fig. S12 for data up to 7 T), consistent with experiments on bulk crystals. It can be seen from the magnetisation loop shown in Figure \ref{fig:magnetometry}b that in-plane zero magnetic field magnetisation is higher than that of the out-of-plane configuration, which is although weaker than in bulk systems, is in agreement with the expected in-plane anisotropy. Zero field magnetisation is found to be $\mathrm{(4.5\pm0.1)\, emu/cm^{3}}$ and the magnetisation offset of linear susceptibility is found to be $\mathrm{(10.5\pm0.5) \,emu/cm^{3}}$. These values expressed as magnetisation per manganese atom are found to be $\mathrm{(0.0086 \pm 0.0002) \,\mu_{B}/Mn}$ and $\mathrm{(0.020 \pm 0.001) \,\mu_{B}/Mn}$ where the former is in agreement with measured value $\approx$ $\mathrm{0.007 \,\mu_{B}/Mn}$ in bulk systems \cite{Yamada1988}. In order to evidence the antiferromagnetic nature of our films further, we have measured a magnetic field polarity-dependent exchange bias effect as shown in \cite{supp}, Figure S13, with a measured bias field of 8 mT. We observed this exchange bias effect without an additional magnetic layer, consistent with experiments on bulk crystals \cite{Qian2014}. The exchange bias originates from the $\text{Mn}_3\text{Ge}$ film as the diffused $\text{Mn}_3\text{Ge}$ into Ru/$\text{Al}_2\text{O}_3$ interface is magnetically coupled to the main film, which we will discuss further in a following work.

\section{\label{sec:conclusion}CONCLUSION}
In conclusion, we have demonstrated optimised growth of non-collinear antiferromagnet \hexagonal $\text{Mn}_3\text{Ge}$(0001) films through utilisation of a single $\text{Mn}_5\text{Ge}$ target, in contrast to previous studies that have reported mixed phase and orientation thin films \cite{Wang2022, Wang2025}. We have demonstrated the \hexagonal phase of the films and their exceptional high quality (RMS roughness $<$ 1 nm) through RHEED, XRR, XRD and (4D) STEM. We find that such a remarkable quality of \hexagonal films is enabled through an initial growth of a \tetragonal phase on a Ru(0001) seed layer grown on a c-plane sapphire substrate and a post annealing induced structural transition. Such a structural transition is found to be reverse of the ones reported on \MnGe films grown on STO substrates \cite{Wang2021_2}. We have further characterised both the pre-anneal \tetragonal and and the post anneal \hexagonal phases through magnetometry. Tetragonal films have magnetic characteristics that are expected from a tilted easy axis for (112) texture, which are in agreement with experiments on (001) films \cite{Kurt2012} as well as polycrystalline nature that is in agreement with bulk polycrystalline films \cite{Yamada1988}. For the \hexagonal $\text{Mn}_3\text{Ge}$(0001) films, we find evidence of the expected non-collinear antiferromagnetic ordering, which manifests itself with an in-plane weak magnetic signal associated with single-ion anisotropy-induced deviations from the perfect $120\,\unit{\degree}$ spin structure. The obtained Néel temperature for the non-collinear antiferromagnetic order onset is comparable but smaller than bulk \hexagonal \MnGe studies with measured magnetisation in contrast being larger \cite{Yamada1988, Nayak2016}. In contrast to other thin-film studies on related non-collinear antiferromagnets, we report an absence of magnetic saturation, suggesting truly preserved non-collinear antiferromagnetic order. Our work suggests single-target growth as a time-efficient method to produce high-quality non-collinear antiferromagnetic $\text{Mn}_3\text{Ge}$(0001) films, where spin Hall effect signals are expected to be maximised \cite{Zhang2017}. 
\section{\label{sec:conclusion}ACKNOWLEDGEMENTS}
We thank the Engineering and Physical Sciences Research Council in the UK through the grant EP/X027074/1. S.O. acknowdleges the support of Henry Royce Institute for Advanced Materials for enabling access to Royce Deposition System facilities at University of Leeds. The STEM work was made possible by the EPSRC Cryo-Enabled Multi-microscopy for Nanoscale Analysis in the Engineering and Physical Sciences (EP/V007661/1). M.S.C. acknowledges funding from the Royal Society Tata University Research Fellowship (URF/R1/201318), Royal Society Enhancement Award (RF/ERE/210200EM1), ERC Consolidator Grant DISCO (101171966), and EPSRC grant EP/V001914/1.

\bibliography{references}

\clearpage

\end{document}


\title{Optimised growth of epitaxial $\mathrm{D0_{19}}$ $\mathrm{Mn_{3}Ge}$ thin films via single target sputtering}

\author{Servet Ozdemir}
\email{s.ozdemir@leeds.ac.uk}
\affiliation{School of Physics and Astronomy, University of Leeds, LS2 9JT, Leeds, UK}

\author{Ping-Luen Ho}
\affiliation{Department of Materials, Imperial College London, London, SW7 2AZ, UK}

\author{Callum Brennan-Rich}
\author{Matthew Rogers}
\affiliation{School of Physics and Astronomy, University of Leeds, LS2 9JT, Leeds, UK}

\author{Philippa Shepley}
\author{Ahmet Yagmur}
\author{Ismet Gelen}
\author{Sehwan Song}
\affiliation{School of Physics and Astronomy, University of Leeds, LS2 9JT, Leeds, UK}

\author{Zabeada Aslam}
\affiliation{School of Chemical and Process Engineering, University of Leeds, LS2 9JT, Leeds, UK}

\author{Mannan Ali}
\affiliation{School of Physics and Astronomy, University of Leeds, LS2 9JT, Leeds, UK}

\author{Shelly Conroy}
\author{Sandrine Heutz}
\affiliation{Department of Materials, Imperial College London, London, SW7 2AZ, UK}

\author{Gavin Burnell}
\email{G.Burnell@leeds.ac.uk}
\author{B J Hickey}
\email{B.J.Hickey@leeds.ac.uk}
\affiliation{School of Physics and Astronomy, University of Leeds, LS2 9JT, Leeds, UK}

\maketitle


\renewcommand{\thefigure}{S\arabic{figure}}
\renewcommand{\thetable}{S\arabic{table}}
\section{Supplementary Figures}
\label{sec:single_col_fig}

\begin{figure}[htbp!]
\includegraphics[scale=0.4]{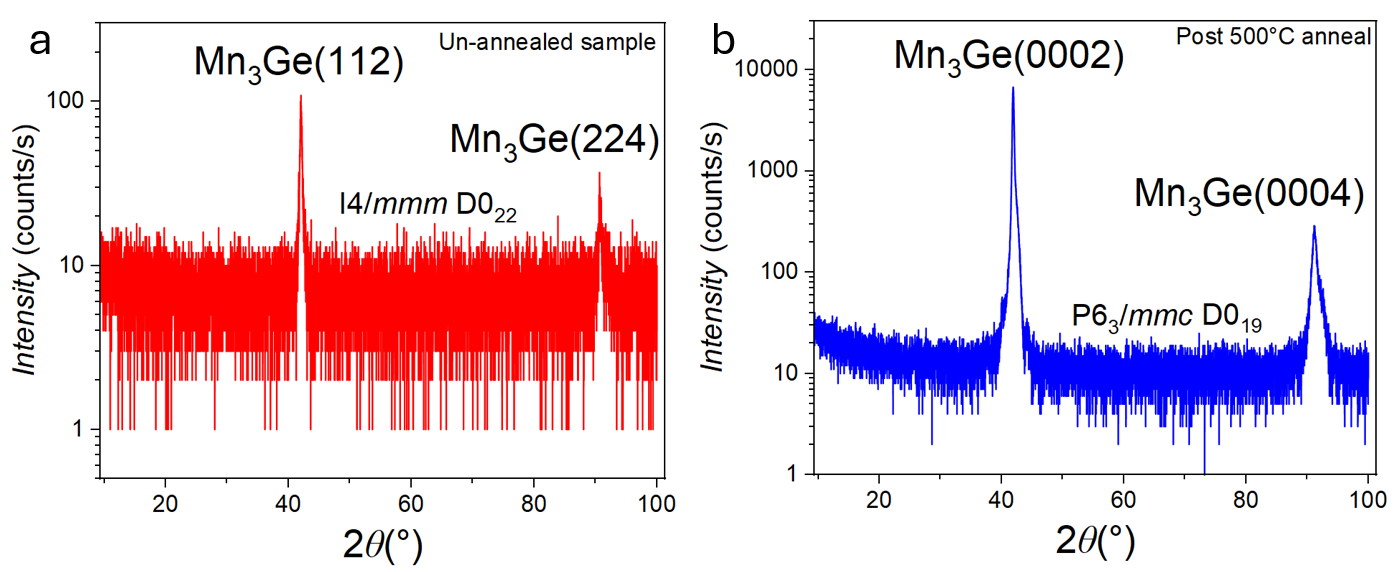}
\centering
\caption{X-ray diffraction curves measured as a function of $\mathrm{2\theta}$ on a, non annealed as grown sample of tetragonal I4/\textit{mmm} phase and (112) texture, and \SI{500}{\celsius} annealed thin film of hexagonal $P6_{3}$/\textit{mmc} phase and (0001) texture. The absence of other prominent peaks suggests no other phases, consistent with 4D-STEM.}
\label{fig:S1}
\end{figure}

\begin{figure}[htbp!]
\includegraphics[scale=0.4]{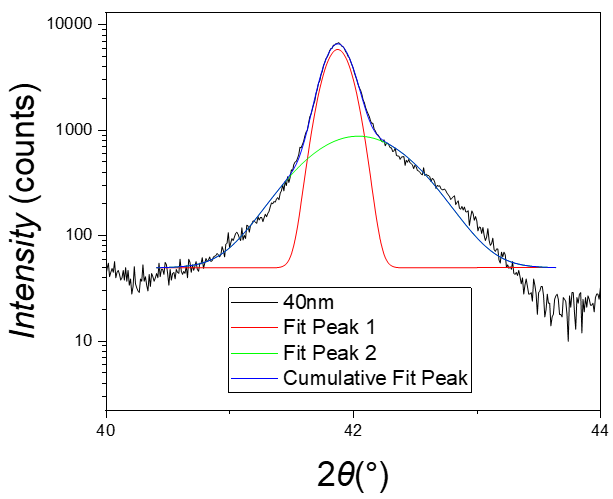}
\centering
\caption{X-ray diffraction peak measured for Ru(0001)/$\text{Mn}_3\text{Ge}$(0001) heterostructure on c-plane sapphire. Two respective overlapping peaks can be seen with (0002) peak for $\text{Mn}_3\text{Ge}$ being at \ang{41.87} and (0002) peak for Ru(0001) being at \ang{42.04}. A c-plane crystallite size of (40.4±0.2) nm is confirmed using Debye-Scherrer formula for the $\text{Mn}_3\text{Ge}$ layer, in agreement with the film thickness, as expected for columnar growth.}
\label{fig:S1}
\end{figure}

\begin{figure}[htbp!]
\includegraphics[scale=0.45]{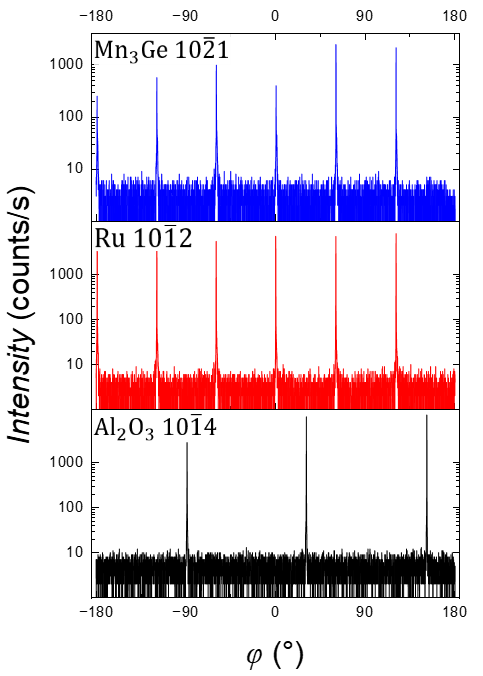}
\centering
\caption{$\phi$-scans carried out probing the planes of (10$\bar{1}$4) for $\text{Al}_2\text{O}_3$, (10$\bar{1}$2) for Ru and (10$\bar{2}$1) for $\text{Mn}_3\text{Ge}$. The apparent 6 peaks for Ru and $\text{Mn}_3\text{Ge}$ arise due to 6-fold anisotropy and suggest that the likely domains arising due to columnar growth have a phase relationship.}
\label{fig:S2}
\end{figure}

\begin{figure}[htbp!]
\centering
\includegraphics[width=\columnwidth]{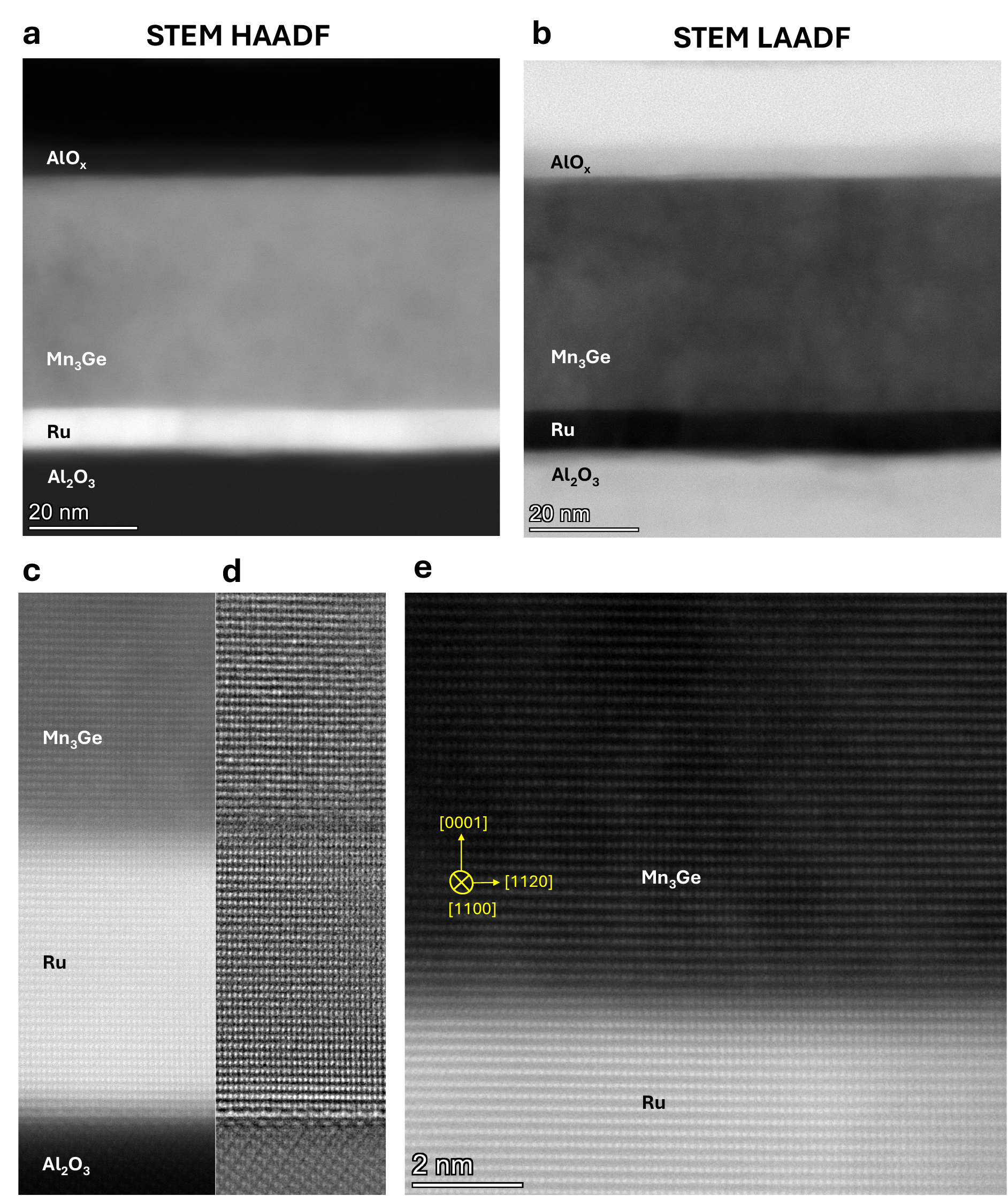}
\caption{Cross-sectional STEM characterisation of the $\text{Mn}_3\text{Ge}$ film annealed at \SI{500}{\celsius}. (a) HAADF-STEM and (b) LAADF-STEM images showing the $\text{Mn}_3\text{Ge}$/Ru/$\text{Al}_2\text{O}_3$ heterostructure with an $\text{AlO}_X$ capping layer. (c) Higher-magnification HAADF-STEM image of the interfaces. (d) Corresponding contrast-enhanced image. (e) Atomic-resolution STEM image showing the hexagonal $\text{D}\text{0}_{19}$ $\text{Mn}_3\text{Ge}$/Ru interface and crystallographic orientation.}
\label{fig:S4}
\end{figure}

\begin{figure}[htbp!]
\includegraphics[scale=0.5]{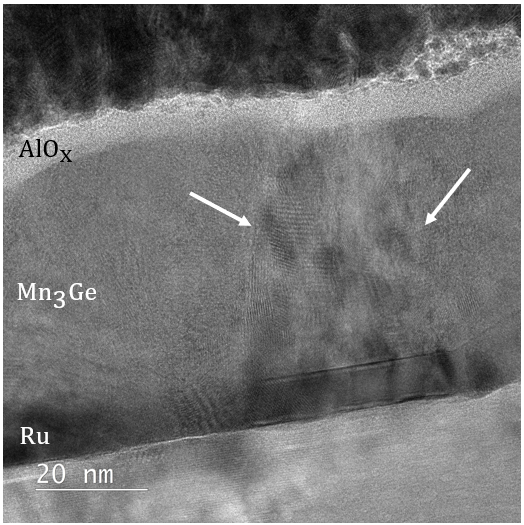}
\centering
\caption{HR-TEM image taken on the non-annealed sample, where the polycrystalline regions can be seen pointed by the arrows. }
\label{fig:S3}
\end{figure}
\begin{figure}[htbp!]
\includegraphics[scale=0.5]{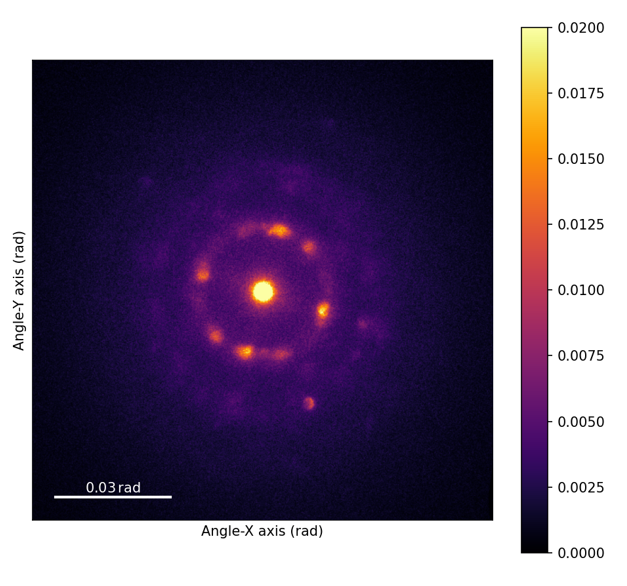}
\centering
\caption{4D-STEM diffraction pattern from the non-annealed $\text{Mn}_3\text{Ge}$ sample, exhibiting rings that suggest polycrystalline regions.}
\label{fig:S3}
\end{figure}

\begin{figure}[htbp!]
\centering
\includegraphics[width=\columnwidth]{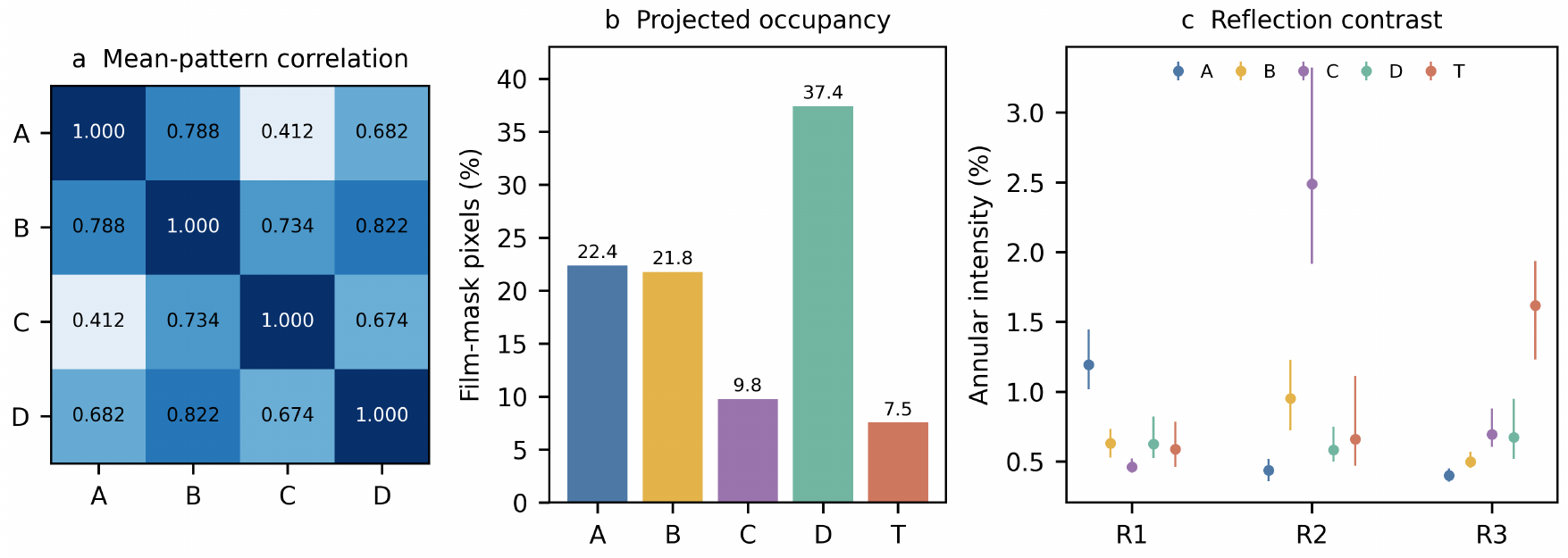}
\caption{Analysis of the 4D-STEM diffraction-pattern classification for the as-grown $\text{Mn}_3\text{Ge}$ film deposited at \SI{250}{\celsius}. (a) Correlation matrix comparing the mean diffraction patterns of classes A--D. (b) Projected occupancy of classes A--D and T within the film mask, showing the percentage of scan positions assigned to each class. (c) Relative intensities of three selected diffraction reflections (R1--R3) for each class, showing differences in diffraction contrast between the classified regions. The dataset contains 6,800 scan positions across a $598.51 \times 101.75$ nm field of view. The broad film mask contains 2,800 positions (rows 9--22), while the central mask contains 2,000 positions (rows 11--20). All row indices are zero-based.}
\label{fig:S6}
\end{figure}

\begin{figure}[htbp!]
\centering
\includegraphics[width=\columnwidth]{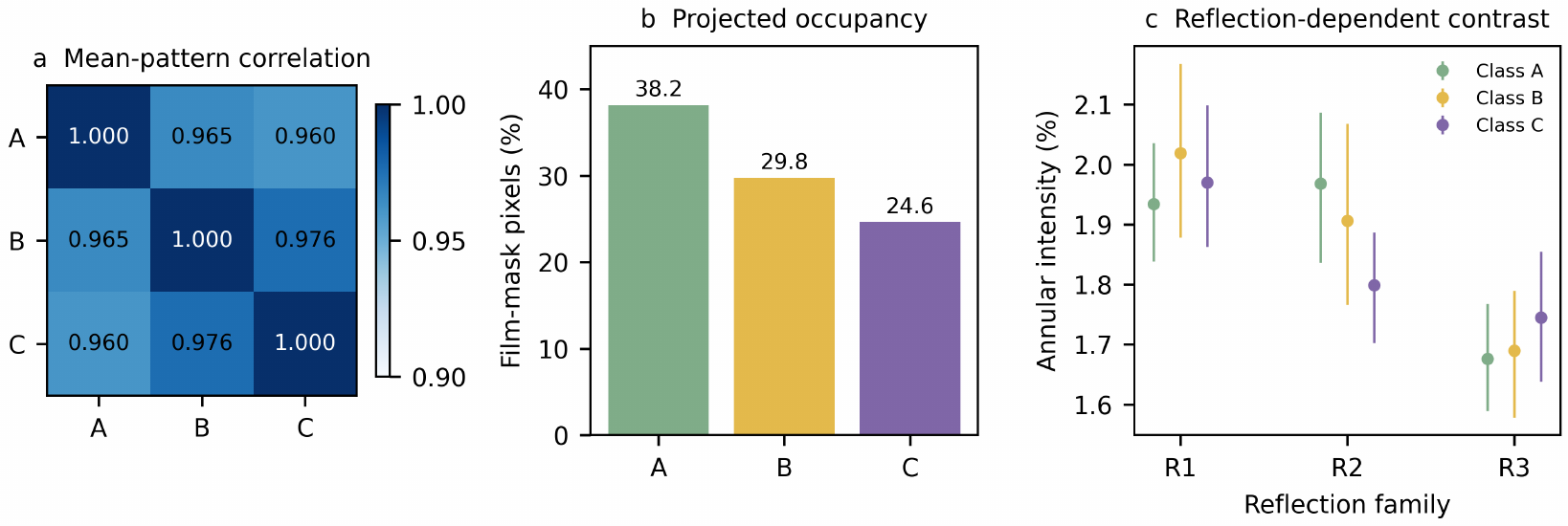}
\caption{Analysis of the 4D-STEM diffraction-pattern classification for the $\text{Mn}_3\text{Ge}$ film annealed at \SI{500}{\celsius}. (a) Pearson correlation matrix comparing the mean diffraction patterns of classes A--C, showing high correlations between the classified regions. (b) Projected occupancy of classes A--C within the film mask, showing the percentage of scan positions assigned to each class. (c) Reflection-dependent contrast for three selected diffraction reflections (R1--R3), showing median intensities with interquartile ranges. The dataset contains $28 \times 200$ scan positions with 2.9925 nm sampling. The broad film mask contains 2,800 positions (rows 7--20), while the central mask contains 2,200 positions (rows 9--19). All indices are zero-based.}
\label{fig:S7}
\end{figure}
\begin{figure}[htbp!]
\includegraphics[scale=0.65]{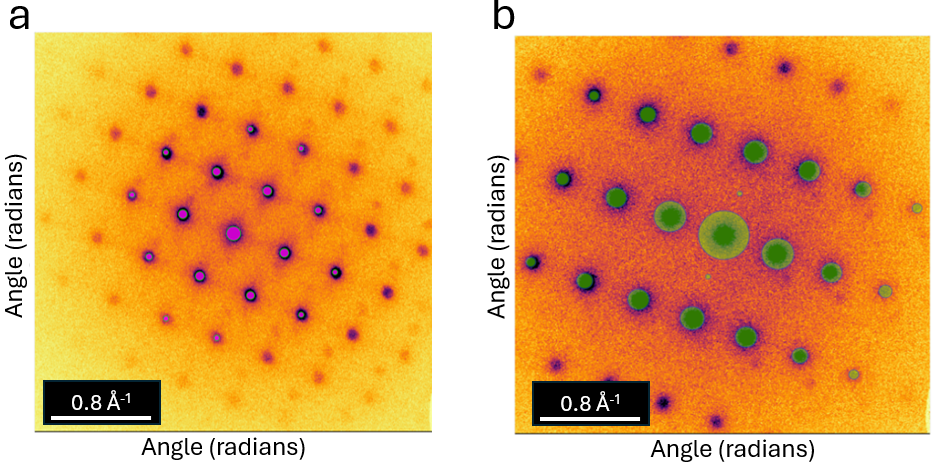}
\centering
\caption{(a) 4D-STEM electron diffraction pattern prominent on a non-annealed sample with an overlaid simulation of tetragonal $\text{D}\text{0}_{22}$ $\text{Mn}_3\text{Ge}$ film with unit cell parameters suggested by RHEED and XRD. (b) 4D-STEM electron diffraction pattern prominent on an annealed sample with an overlaid simulation of hexagonal $\text{D}\text{0}_{19}$ $\text{Mn}_3\text{Ge}$ film with unit cell parameters suggested by RHEED and XRD.}
\label{fig:S2}
\end{figure}

\begin{figure}[htbp!]
\includegraphics[scale=0.5]{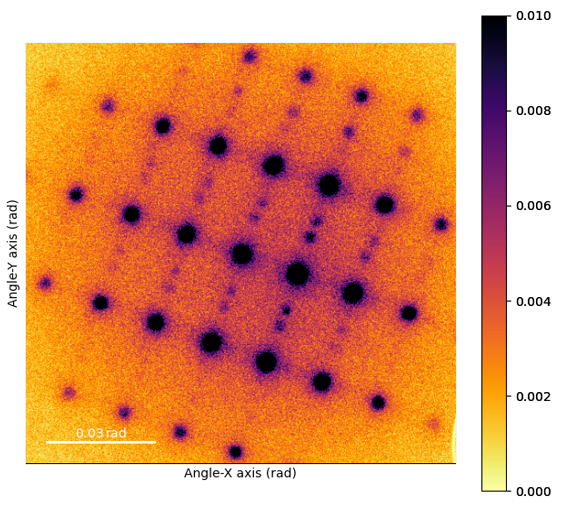}
\centering
\caption{4D-STEM diffraction pattern measured on the \SI{500}{\celsius} annealed hexagonal $\text{D}\text{0}_{19}$ $\text{Mn}_3\text{Ge}$ sample, where unevenly spaced smaller spots are visible on top of the diffraction pattern reported in main text, that we attribute to formation of an incommensurate super-lattice on lamella surface.}
\label{fig:S3}
\end{figure}

\begin{figure}[htbp!]
\centering
\includegraphics[scale=0.95]{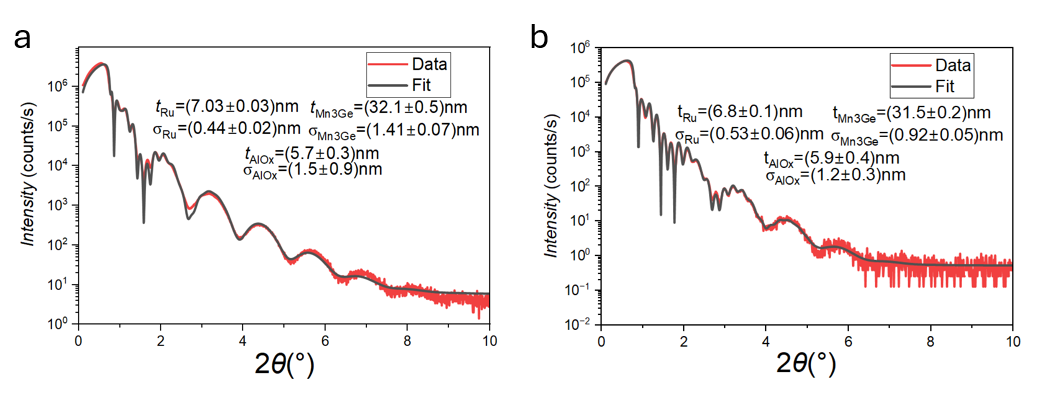}
\caption{X-ray reflectivity (XRR) measurements of $\text{Mn}_3\text{Ge}$ thin films before and after annealing. (a) XRR measurement and corresponding fit for the as-grown film deposited at \SI{250}{\celsius}, giving a $\text{Mn}_3\text{Ge}$ thickness of $32.1 \pm 0.5$ nm and $\text{Mn}_3\text{Ge}$ RMS roughness of $1.41 \pm 0.07$ nm. (b) XRR measurement and corresponding fit following annealing at \SI{500}{\celsius}, showing a $\text{Mn}_3\text{Ge}$ thickness of $31.5 \pm 0.2$ nm and reduced $\text{Mn}_3\text{Ge}$ RMS roughness of $0.92 \pm 0.05$ nm. The fitted results indicate that RMS roughness of $\text{Mn}_3\text{Ge}$ decreases after annealing.}
\label{fig:S5}
\end{figure}

\begin{figure}[htbp!]
\includegraphics[scale=0.5]{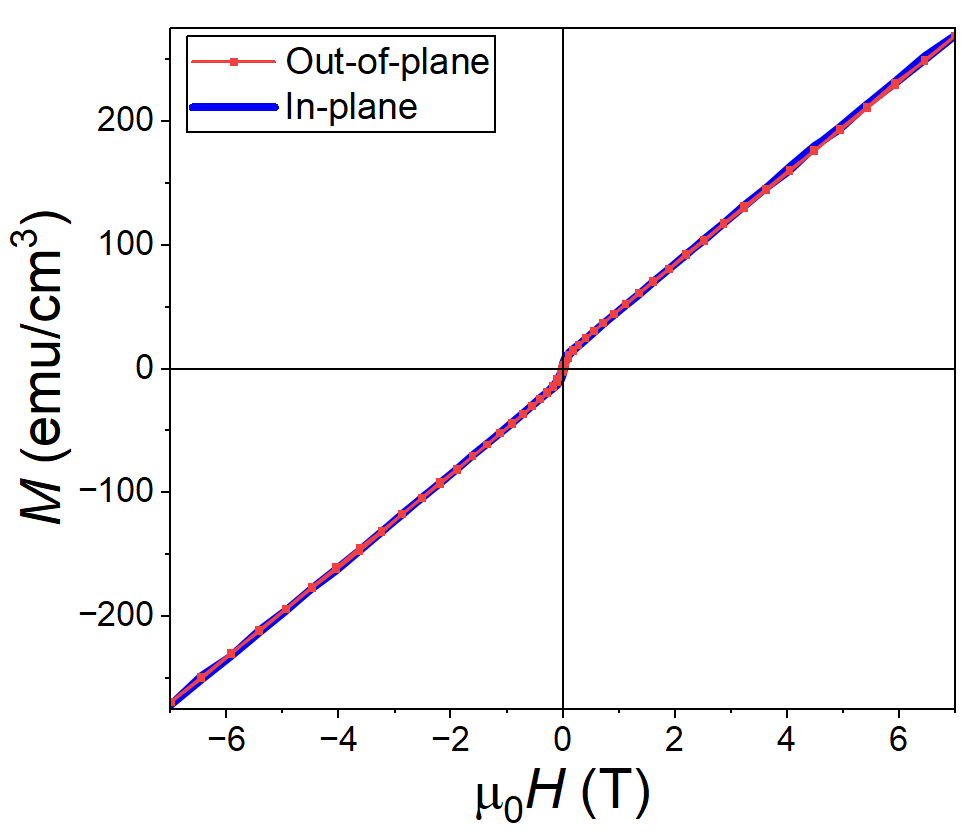}
\centering
\caption{In-plane and out-of-plane magnetic hysteresis loops measured up to 7 T on a 40 nm $\text{D}\text{0}_{19}$ $\text{Mn}_3\text{Ge}$ sample showing absence of saturation at \textit{T} = 10 K.}
\label{fig:S3}
\end{figure}
\begin{figure}[htbp!]
\includegraphics[scale=0.5]{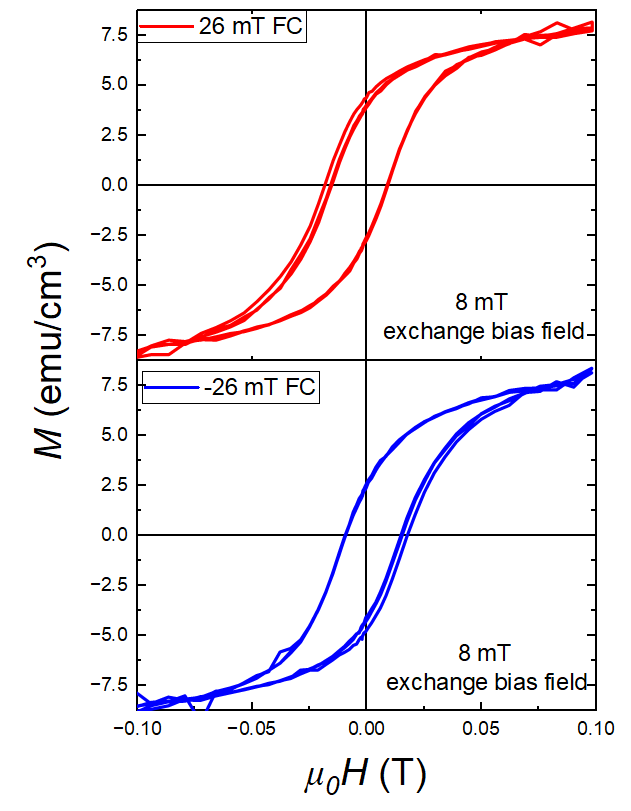}
\centering
\caption{Positive (top panel) and negative (bottom panel) field cooled exchange bias effect on a $\mathrm{50\,nm}$ hexagonal $\text{D}\text{0}_{19}$ sample measured at \textit{T} = 10 K. The exchange bias can't be attributed to Mn$_{3}$Ge layer that has diffused into Ru/$\text{Al}_2\text{O}_3$ interface, this is because the system is behaving as a single magnetic system, which we will discuss further in a future study.}
\label{fig:S3}
\end{figure}
\clearpage
\begin{table}[htbp]
    \centering
    \includegraphics[width=0.85\textwidth]{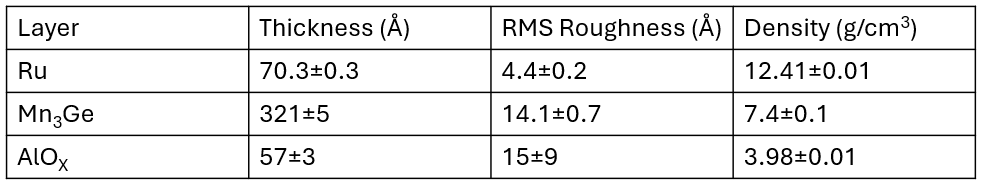} 
    \caption{Un-annealed sample X-ray reflectivity fitting parameters.}
    \label{tab:tableT1}
\end{table}

\begin{table}[htbp]
    \centering
    \includegraphics[width=0.85\textwidth]{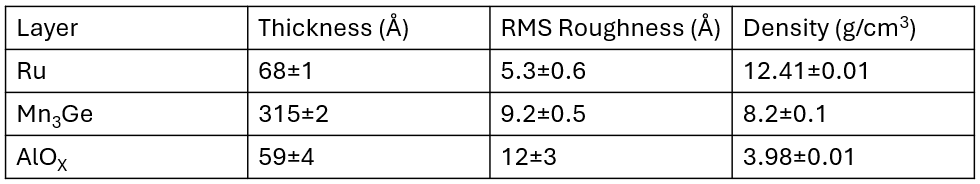}
    \caption{\SI{500}{\celsius} annealed sample x-ray reflectivity fitting parameters.}
    \label{tab:tableT2}
\end{table}

\section{Supplementary Note}
In order to calculate the angle between the (112) plane normal, $\hat{\boldsymbol{n}_{i}}$, and the c axis [001], we can use the dot product formula

\begin{equation}
\begin{split}
\cos{(\theta)} = \frac{\hat{\boldsymbol{n}_{i}}\cdot[001]}{\left|{\hat{\boldsymbol{n}_{i}}}\right|\left|[001]\right|} 
\end{split}
\label{eq:three}
\end{equation}
Taking $\hat{\boldsymbol{n}_{i}} = \frac{1}{a}\hat{\boldsymbol{i}}+\frac{1}{a}\hat{\boldsymbol{j}}+\frac{2}{c}\hat{\boldsymbol{k}}$ and [001] = c$\hat{\boldsymbol{k}}$ we obtain the expression for $\cos{(\theta)}$ as
\begin{equation}
\begin{split}
\cos{(\theta)} = \frac{1}{\sqrt{\frac{1}{2}+(\frac{a}{c})^2}} 
\end{split}
\label{eq:three}
\end{equation}
We have determined the ratio ($\frac{a}{c}$) to be 0.52 which yields the angle $\theta$ to be 58.4\unit{\degree}.